\documentclass[runningheads,a4paper]{llncs}

\usepackage{amssymb}
\usepackage{amsmath}
\usepackage{bm}
\usepackage{graphicx}
\usepackage{color}
\usepackage{hyperref}

\usepackage{url}

\begin{document}

\mainmatter  

\title{Myocardial Segmentation of Late Gadolinium Enhanced MR Images by Propagation of Contours from Cine MR Images}

\titlerunning{Myocardial Segmentation of Late Gadolinium Enhanced MR Images}

%
%
\author{Dong Wei\inst{1} \and
Ying Sun\inst{1} \and
Ping Chai \inst{2} \and
Adrian Low\inst{2} \and
Sim Heng Ong\inst{1}}
\authorrunning{D. Wei et al.}

\institute{Department of Electrical and Computer Engineering,\\
National University of Singapore, Singapore\\
\email{\{dongwei, elesuny, eleongsh\}@nus.edu.sg}
\and
Cardiac Department, National University Heart Centre, Singapore}

%
%

\toctitle{Myocardial Segmentation of Late Gadolinium Enhanced MR Images by Propagation of Contours from Cine MR Images}
\maketitle

\begin{abstract}
Automatic segmentation of myocardium in Late Gadolinium Enhanced (LGE) Cardiac MR (CMR) images is often difficult due to the intensity heterogeneity resulting from accumulation of contrast agent in infarcted areas.
In this paper, we propose an automatic segmentation framework that fully utilizes shared information between corresponding cine and LGE images of a same patient.
Given myocardial contours in cine CMR images, the proposed framework achieves accurate segmentation of LGE CMR images in a coarse-to-fine manner.
Affine registration is first performed between the corresponding cine and LGE image pair, followed by nonrigid registration, and finally local deformation of myocardial contours driven by forces derived from local features of the LGE image.
Experimental results on real patient data with expert outlined ground truth show that the proposed framework can generate accurate and reliable results for myocardial segmentation of LGE CMR images.
\end{abstract}

\section{Introduction}
\label{sec:introduction}

Viability assessment of the myocardium after the experience of a myocardial infarction is essential for diagnosis and therapy planning.
In particular, the detection, localization and quantification of infarcted myocardial tissue, also called infarct/infarction/scar,
are important for determining whether and which part(s) of a heart that has undergone a myocardial infarction may benefit from re-vascularization therapy.
In a typical CMR examination, a contrast agent is injected and a cine sequence is acquired approximately at the same time;
15 to 20 minutes later, an LGE scan is performed, and by then scars exhibit brighter (enhanced) intensities than healthy myocardium.
This is because scars accumulate more contrast agent and experience delayed wash-in and wash-out of contrast agent.
Typical LGE images with enhanced scars can be seen in Fig.~\ref{fig:segmentationResults}.

Delineation of myocardial contours is the first step in locating and quantifying infarctions in LGE images.
Because manual delineation is not only time-consuming but also subject to  intra- and inter-observer variability, it is highly desirable to automate the contour delineation process.
However, automatic segmentation of myocardial contours is a difficult task, due mainly to the intensity inhomogeneity of the myocardium resulting from the accumulation of contrast agent in infarcted areas.
To the best of our knowledge, there has been little research aimed at fully automatic myocardial segmentation in LGE images, and there is no commercially or publicly available automatic segmentation tool for clinical use.
Most of the existing approaches utilize pre-delineated myocardial contours in the corresponding cine MRI as \emph{a priori} knowledge~\cite{Ciofolo2008,Dikici2004}.
Such an approach is reasonable because the patient is asked to stay still during the entire acquisition process and there are many methods available for automatic segmentation of cine MRI~\cite{hautvast2006automatic}.
Nevertheless, major difficulties in this approach include:
(i) misalignment and nonrigid deformation between cine and LGE data due to respiratory motion and/or imperfectness of electrocardiography gating
(which will be worse in cases of arrhythmia); and
(ii) differences in resolution, field of view, and global intensity of cine and LGE data.

In this paper, we propose a multi-level framework for automatic myocardial segmentation of LGE CMR images in a coarse-to-fine manner.
Our work is different from the aforementioned ones~\cite{Ciofolo2008,Dikici2004} in three respects:
(i) we fully utilize shared information between corresponding cine and LGE images by registering the cine image to the LGE image in an affine-to-nonrigid manner, including both shape and intensity information;
(ii) instead of using conventional similarity metrics such as mutual information and cross-correlation, experimentally we choose pattern intensity~\cite{PI1997} which leads to accurate nonrigid registration of corresponding cine and LGE images;
and (iii) at the finest level of segmentation, we propose to detect endocardial edges by adaptively selecting one of the two cases: normal endocardium and sub-endocardial scars, as well as, incorporate a new effective thickness constraint into the evolution scheme based on the simplex mesh~\cite{simplexmeshIJCV1999} geometry.
The rest of the paper is organized as follows.
Section~\ref{sec:method} describes the proposed automatic segmentation framework.
Section~\ref{sec:results} presents the experimental results on real patient data, followed by the conclusion in Section~\ref{sec:conclusion}.

\section{Method}
\label{sec:method}

\begin{figure}
  \centering
  \includegraphics[width=0.95\textwidth]{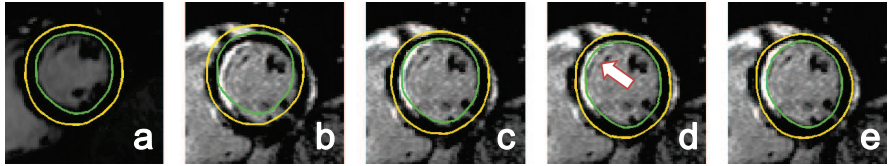}\\
  \caption{The segmentation results obtained at the different stages of our framework: (a) the \emph{a priori} segmentation in the cine image; (b) direct overlay of the \emph{a priori} onto the LGE image without any processing; (c) after affine transformation; (d) after b-spline based FFD; (e) after local deformation of contours. Note that from (d) to (e) a small defect in the \emph{a priori} is corrected (as highlighted by the arrow).}\label{fig:procedures}
\end{figure}

Our automatic segmentation framework comprises three major steps from coarse to fine levels:
(i) estimate an affine transformation by maximizing normalized cross-correlation~(NCC) for rough alignment of corresponding cine and LGE images;
(ii) nonrigidly register the affine-transformed cine image to the LGE image using B-spline based free-form deformation (FFD)~\cite{Rueckert1999} with pattern intensity~\cite{PI1997} as the similarity metric;
and (iii) further deform the myocardial contours obtained from the previous step based on local features of the LGE image using the simplex mesh geometry~\cite{simplexmeshIJCV1999}.
Figure~\ref{fig:procedures} shows the segmentation results obtained at the different stages of our framework.

\subsection{Pre-Processing and Affine Transformation}

In this work, we only consider short-axis slices.
A cine sequence comprises multiple slices at different phases, covering a complete cardiac cycle.
Unlike cine data, a set of LGE data comprises slices at only one phase, which is often located between end-systole and end-diastole.
Therefore, given a target LGE image, the first task is to select a corresponding cine image.
We select the cine image with the same spatial location and the closest phase to the LGE image according to their respective DICOM header information, and delineate myocardial contours in the selected cine image as \emph{a priori} segmentation.
Then we normalize each pair of corresponding cine and LGE images to:
(i)   the same physical resolution;
(ii)  similar histogram distributions (specifically, we specify the histogram of the LGE image to resemble that of the cine image); and
(iii) the same size.

The coarse level segmentation is achieved by registering the cine image to the target LGE image through a constrained affine transformation.
For the 2D case the affine transformation matrix in homogeneous coordinates can be expressed as
\begin{equation}
    \bm{A}=\begin{bmatrix}
            a_{11}\quad & a_{12}\quad & a_{13} \\
            a_{21}\quad & a_{22}\quad & a_{23} \\
            0\quad & 0\quad & 1 \\
            \end{bmatrix}.
\end{equation}
Based on the assumption that there should not be any significant rotation or shearing effects, and the scaling and translation effects should also be small, we constrain the estimated affine transformation so that:
\begin{equation}\label{eq:affineconstraint}
    |a_{11}-1|,\ |a_{22}-1|<\varepsilon_{\mathrm{scale}}\quad\mbox{and}\quad |a_{13}|,\ |a_{23}|<\varepsilon_{\mathrm{translate}},
\end{equation}
where $\varepsilon_{\mathrm{scale}}$ and $\varepsilon_{\mathrm{translate}}$ are corresponding thresholds and set to 0.1 and 10 for our data.
Here we use NCC as the similarity measure because it is invariant to both shift and scale intensity distortions~\cite{eccPAMI2008} and hence can overcome systematic intensity variations  between corresponding cine and LGE images.
For a fast implementation, we adopt the enhanced cross-correlation (ECC) algorithm proposed in~\cite{eccPAMI2008}, which optimizes an efficient approximation of NCC and leads to a closed-form solution in each iteration.

\subsection{B-Spline based Nonrigid Registration}
\label{ssec:spline}

Since the deformation of a heart cannot be completely described by an affine transformation, we apply B-spline based nonrigid registration~\cite{Rueckert1999} following the affine transformation.
For a 2D image, B-spline based FFD is controlled by a mesh of control points $\Phi=\{\phi_{i,j}\}$.
Let $s_{x}\times s_{y}$ denote the uniform spacing of the mesh grid, then the FFD can be written as:
\begin{equation}\label{eq:B-spline}
    \bm{T}(x,y,\Phi)=\sum_{l=0}^{3}\sum_{m=0}^{3}B_{l}(u)B_{m}(v)\phi_{i+l,j+m},
\end{equation}
where $i=\lfloor x/s_{x}\rfloor-1$, $j=\lfloor y/s_{y}\rfloor-1$, $u=x/s_{x}-\lfloor x/s_{x}\rfloor$, $v=y/s_{y}-\lfloor y/s_{y}\rfloor$, and $B_{l}$ is the $l$th basis function of B-splines.
The transformation field $\bm{T}$ completely determines deformation of the whole image.
Because B-splines are locally controlled, they are relatively more efficient compared to other kinds of splines such as thin-plate splines, even for a large number of control points.

Different from most registration methods which use conventional similarity measures such as normalized mutual information (NMI), experimentally we select \emph{pattern intensity}~\cite{PI1997} as the similarity measure for the nonrigid registration.
Given two images $I_{1}$ and $I_{2}$, pattern intensity operates on the difference image
$I_{\mathrm{diff}}=I_{1}-I_{2}$.
If $I_{1}$ and $I_{2}$ are two images of the same object and well registered, structures from this object should vanish and there should be a minimum number of structures or patterns in $I_{\mathrm{diff}}$.
A suitable similarity measure should, therefore, characterize the \emph{structuredness} of $I_{\mathrm{diff}}$.
Pattern intensity considers a pixel of $I_{\mathrm{diff}}$ to belong to a structure if it has a significantly different value from its neighboring pixels.
Using a constant radius $r$ to define the neighborhood of the pixel being examined, pattern intensity is defined as:
\begin{equation}\label{eq:PI}
\begin{split}
  P_{r,\sigma}(I_{1},I_{2})=&P_{r,\sigma}(I_{\mathrm{diff}}) \\
                           =&\frac{1}{N_{I_{\mathrm{diff}}}}\sum_{x,y}\frac{1}{N_{r}}\sum_{(x-v)^2+(y-w)^2\leq r^2}\frac{\sigma^2}{\sigma^2+[I_{\mathrm{diff}}(x,y)-I_{\mathrm{diff}}(v,w)]^2},
\end{split}
\end{equation}
where $N_{I_{\mathrm{diff}}}$ and $N_{r}$ denote the number of pixels in $I_{\mathrm{diff}}$  and the neighborhood respectively.
A constant $\sigma$ is introduced to suppress the impact of noise.

\begin{figure}
  \centering
  \includegraphics[width=0.95\textwidth]{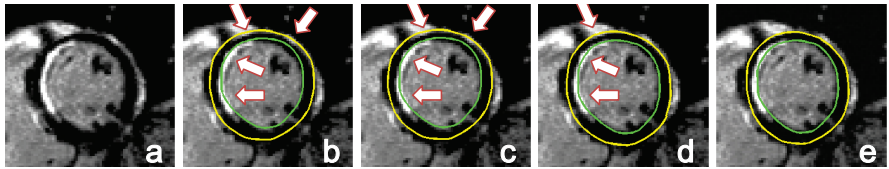}\\
  \caption{A comparison of the effects of different similarity metrics on b-spline based cine and LGE images registration: (a) the LGE image; (b) squared intensity difference; (c) NCC; (d) NMI; (e) pattern intensity. The bright arrows in (b)-(d) highlight the locations where pattern intensity based registration outperforms others.}\label{fig:compareSM}
\end{figure}

Pattern intensity has the following desirable properties:
(i) small deviations in intensity would still produce measurements remaining near to the maximum value due to the introduction of $\sigma$;
(ii) large differences of intensity values have the same effect regardless of their magnitude due to the asymptotic nature, making the measure robust to a few large differences in pixel intensity; and
(iii) its regional nature is able to reduce the effect of differences with scales larger than $r$.
These properties make it robust to the presence of enhanced infarctions of relatively small area in LGE images.
Indeed, we have experimented with popular similarity metrics including squared intensity difference, NMI, and NCC, and found that pattern intensity gives best results for most LGE images in our database (see Fig.~\ref{fig:compareSM} for a demonstration).
This is consistent with the findings from a comparative study of six similarity measures for intensity-based 2D-3D vertebra image registration~\cite{PIcomparison1998}, which state that pattern intensity outperformed the others and was able to achieve accurate registration even when soft-tissue structures and interventional instruments were present as differences between the images.

In our implementation, we use the optimization scheme described in~\cite{Rueckert1999}.
The cost functional to be minimized consists of two terms, one for similarity and the other for smoothness.
The similarity cost is defined as:
\begin{equation}\label{eq:ffdsimilarcost}
    \mathcal{C}_{\mathrm{similar}}[I_{\mathrm{1}},\bm{T}(I_{\mathrm{2}},\Phi)]=1-P_{r,\sigma}[I_{\mathrm{1}},\bm{T}(I_{\mathrm{2}},\Phi)],
\end{equation}
and the smoothness cost $\mathcal{C}_{\mathrm{smooth}}$ is defined as the 2-D bending energy of a thin-plate of metal to regularize $\bm{T}$~\cite{Rueckert1999}.
The final cost functional is a weighted sum of $\mathcal{C}_{\mathrm{similar}}$ and $\mathcal{C}_{\mathrm{smooth}}$ with a weighting factor $\lambda$ which controls the trade-off between accurate match of the two images and smoothness of the transformation:
\begin{equation}\label{eq:ffdfinalcost}
    \mathcal{C}(\Phi)=\mathcal{C}_{\mathrm{similar}}[I_{\mathrm{1}},\bm{T}(I_{\mathrm{2}},\Phi)]+\lambda\mathcal{C}_{\mathrm{smooth}}(\bm{T}).
\end{equation}
We have experimentally determined the values for grid spacing $s_{x}$, $s_{y}$ and weighting factor $\lambda$.
We found that setting $s_{x}$, $s_{y}=8$ and $\lambda=0.2$ would produce relatively reasonable yet smooth segmentation for most LGE images in our database.

\subsection{Local Deformation of Myocardial Contours}
\label{ssec:ffd}

After the first two steps, we can already obtain reasonable segmentation results which are quite close to true myocardial boundaries.
However, some minor flaws still exist, e.g. a fraction of the infarct is excluded from the myocardial contours (Fig.~\ref{fig:procedures}(d)).
Thus we further locally deform the existing myocardial contours to ensure that scars are enclosed by the final contours, and improve segmentation accuracy.
Different from the previous two steps in which segmentation is achieved by registration of corresponding cine and LGE images, in this step we directly deform the contours based on local features of the LGE image alone.

\begin{figure}
  \centering
  \includegraphics[width=0.95\textwidth]{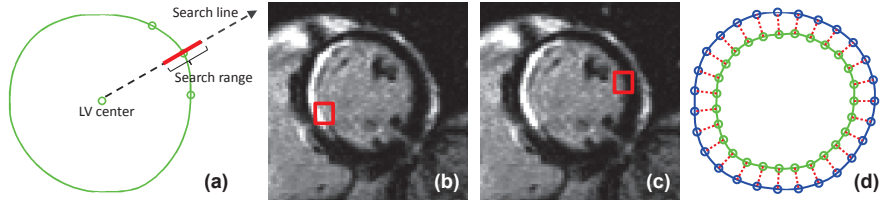}\\
  \caption{An illustration of $\bm{F}_{\mathrm{edge}}$ and $\bm{F}_{\mathrm{thick}}$: (a) the search for myocardial edges along radial directions of the LV; (b) an edge of sub-endocardial infarction highlighted with the red square; (c) an edge of normal endocardium highlighted with the red square; (d) connections established between endo- and epicardial contours as thickness constraint.}\label{fig:FFD}
\end{figure}

We use the \emph{simplex mesh}~\cite{simplexmeshIJCV1999} geometry and a Newtonian mechanical model to represent and deform the contours.
We represent both endo- and epicardial contours with 80 mesh vertices.
At each vertex we define three different forces which move it jointly, namely, smoothness $\bm{F}_{\mathrm{smooth}}$, edge attraction $\bm{F}_{\mathrm{edge}}$ and myocardium thickness $\bm{F}_{\mathrm{thick}}$.
$\bm{F}_{\mathrm{smooth}}$ imposes uniformity of vertex distribution and continuity of simplex angles ~\cite{simplexmeshIJCV1999}.
$\bm{F}_{\mathrm{edge}}$ draws vertices towards detected myocardial edge points along radial directions of the left ventricle (LV).
We detect the edge points by searching pixels along the radial directions in a limited range (Fig.~\ref{fig:FFD}~(a)) and picking those with maximum intensity changes with respect to their neighbors on the search line.
Specially, while searching for endocardial edge points, we consider two cases: edges of sub-endocardial scars (Fig.~ \ref{fig:FFD}~(b)) and normal endocardium (Fig.~\ref{fig:FFD}~(c)).
The former case is made much easier to detect nowadays thanks to the improvements in MRI acquisition techniques
(which improve the contrast between the blood pool and sub-endocardial scars), and given a higher priority over the latter considering the prevalence of sub-endocardial scars in infarction patients. 
In addition, the search range is confined to a narrow band of 4 pixels, due to the closeness of the contours obtained in the previous step to real myocardial boarders, and to avoid an unstable large extent of deformation.
Letting $\hat{p}_{e}$ denote the detected edge point for any vertex $p$, $\bm{F}_{\mathrm{edge}}$ can be expressed as:
\begin{equation}\label{eq:extForce}
    \bm{F}_{\mathrm{edge}}=\omega_{p}(\hat{p}_{e}-p),
\end{equation}
where $\omega_{p}$ is a weight linearly and directly proportional to the local intensity change at $\hat{p}_{e}$ and normalized to the interval $[0,1]$.
Finally, $\bm{F}_{\mathrm{thick}}$ establishes a connection between nearest endo- and epicardial vertices along the radial directions (Fig.~\ref{fig:FFD}(d)), and aims to overcome the problem that arises when the edge search fails (e.g. when there is poor contrast between scars and the blood pool).
Letting $p_{\mathrm{endo}}^{0}$ denote any vertex on the endocardial contour before deformation, and $p_{\mathrm{epi}}^{0}$ its nearest neighbor on the epicardial contour, we introduce $\bm{F}_{\mathrm{thick}}$ as:
\begin{equation}\label{eq:simplexConnectForce}
    \begin{split}
        &\bm{F}_{\mathrm{epi,thick}}=p_{\mathrm{endo}}^{t}+(p_{\mathrm{epi}}^{0}-p_{\mathrm{endo}}^{0})-p_{\mathrm{epi}}^{t},\\
        &\bm{F}_{\mathrm{endo,thick}}=p_{\mathrm{epi}}^{t}-(p_{\mathrm{epi}}^{0}-p_{\mathrm{endo}}^{0})-p_{\mathrm{endo}}^{t},
    \end{split}
\end{equation}
where $p_{\mathrm{\ast}}^{t}$ is the vertex after $t$th iteration.

Letting $p^{t}$ denote the position of any vertex at time $t$, the deformation scheme can be written as
\begin{equation}\label{eq:ffdScheme}
    p^{t+1}=p^{t}+(1-\gamma)(p^{t}-p^{t-1})+\alpha\bm{F}_{\mathrm{smooth}}+\beta\bm{F}_{\mathrm{edge}}+\theta\bm{F}_{\mathrm{thick}},
\end{equation}
where $\gamma$ is a damping factor.
In each iteration, we first fix the endocardial contour and update the epicardial contour, and then update the endocardial contour with the epicardial contour fixed.
Various weights in (\ref{eq:ffdScheme}) are experimentally determined as: $\gamma=0.7$, $\alpha=0.35$, $\beta=0.15$, $\theta=0.1$.

\section{Experimental Results}
\label{sec:results}

\begin{figure}
  \centering
  \includegraphics[width=0.95\textwidth]{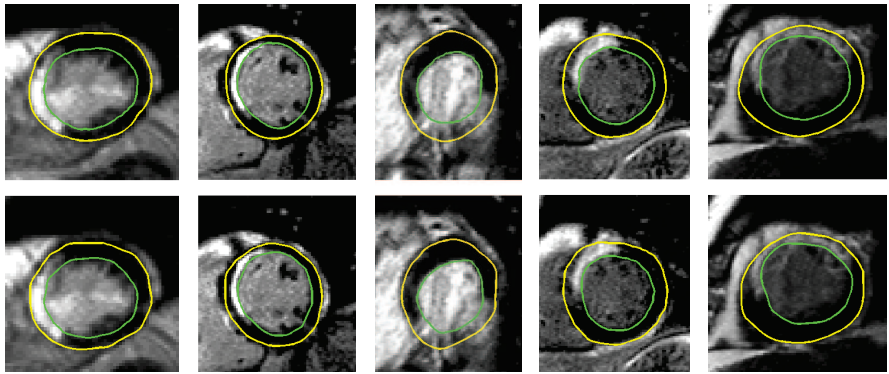}\\
  \caption{Some example segmentation results of our automatic framework (top row), as compared with those by an expert (bottom row).}\label{fig:segmentationResults}
\end{figure}

We have tested the proposed automatic segmentation framework on 59 short-axis LGE CMR images (with corresponding pre-segmented cine CMR images as \emph{a priori} knowledge) from 10 patients that are clinically diagnosed as having experienced myocardial infarction.
The images were acquired with a Siemens Symphony MRI scanner.
Based on visual examination, we find that segmentation results produced by our framework are consistently correct and accurate for nearly all of the LGE images.
Figure~\ref{fig:segmentationResults} shows some example results, together with manual segmentation by an expert.
We have also conducted a quantitative evaluation of our framework by calculating distance errors and the Dice coefficient between the automatic results and the expert's manual segmentation.
The average and maximum errors between the automatic and manual contours are: $0.97\pm0.45$ and $2.60\pm1.15$ pixels for the endocardium, $0.89\pm0.40$ and $2.38\pm0.98$ pixels for the epicardium, and $0.93\pm0.42$ and $2.49\pm1.06$ pixels for both.
This level of error is quite close to intra- and inter-observer variability.
As to the area similarity, the Dice coefficient is $93.59\pm3.80\%$ for the endocardium, $95.63\pm2.62\%$ for the epicardium, and $82.49\pm9.27\%$ for the myocardium.
Since the ultimate goal of automatic myocardial segmentation in LGE CMR images is to serve subsequent automatic quantification of scars, we have further implemented the automatic scar segmentation algorithm proposed in~\cite{tao2010automated} to compare scars within the automatic and manual contours.
The volumetric Dice coefficient between segmented scars within respective contours is $79.78\pm9.72\%$, which is comparable to those reported in \cite{tao2010automated}.

\section{Conclusion}
\label{sec:conclusion}

In this paper we present an automatic segmentation framework for LGE CMR data that fully utilizes shared information between corresponding cine and LGE images.
Given myocardial contours in cine CMR images, our framework first generates intermediate segmentation results by registering the cine image to the target LGE image in an affine-to-nonrigid manner, and then obtains final segmentation results by locally deforming the myocardial contours with forces derived from local features of the LGE image alone.
For the nonrigid registration, different from most other registration methods, we experimentally chose pattern intensity as the similarity metric; it works robustly for most of the LGE images whereas popular similarity metrics such as NMI often cannot compare.
For the local deformation of the myocardial contours, we use the simplex mesh geometry to represent the contours and displace each vertex according to the combination of three different forces: smoothness, edge attraction, and thickness.
Our experimental results on real patient data, along with the quantitative evaluation, show that the proposed automatic segmentation framework can successfully segment short-axis LGE images with good accuracy and reliability.

\bibliographystyle{splncs03}
\bibliography{miccai-782}

\end{document}